\documentclass[a4paper]{article}

\usepackage{INTERSPEECH2021}

\usepackage{amsmath,graphicx}
\usepackage{xcolor,array}
\usepackage{tikz}
\usepackage{nicefrac}

\usetikzlibrary{dsp}
\usetikzlibrary{chains}

\title{Y$^2$-Net FCRN for Acoustic Echo and Noise Suppression}
\name{Ernst Seidel, Jan Franzen, Maximilian Strake, Tim Fingscheidt}
\address{
  Institute for Communications Technology, Technische Universität Braunschweig, \\
	Schleinitzstr. 22, 38106 Braunschweig, Germany}
\email{ \{e.seidel, j.franzen, m.strake, t.fingscheidt\}@tu-bs.de }

\begin{document}

\maketitle
\begin{abstract}
	In recent years, deep neural networks (DNNs) were studied as an alternative to traditional acoustic echo cancellation (AEC) algorithms. The proposed models achieved remarkable performance for the separate tasks of AEC and residual echo suppression (RES). A promising network topology is a fully convolutional recurrent network (FCRN) structure, which has already proven its performance on both noise suppression and AEC tasks, individually. However, the combination of AEC, postfiltering, and noise suppression to a single network typically leads to a noticeable decline in the quality of the near-end speech component due to the lack of a separate loss for echo estimation. In this paper, we propose a two-stage model \mbox{(Y$^2$-Net)} which consists of two FCRNs, each with two inputs and one output (Y-Net). The first stage (AEC) yields an echo estimate, which---as a novelty for a DNN AEC model\textemdash is further used by the second stage to perform RES and noise suppression. While the subjective listening test of the Interspeech 2021 AEC Challenge mostly yielded results close to the baseline, the proposed method
	scored an average improvement of $0.46$ points over the baseline on the blind testset in double-talk on the instrumental metric DECMOS, provided by the challenge organizers. 
	
\end{abstract}

\section{Introduction}

Acoustic echoes arise if a speech communication system's microphone picks up again the echo that was just played by the system's loudspeaker in so-called handsfree mode. If not suppressed, the far-end (FE) speaker is forced to listen to his or her own echo, which considerably lowers the quality of the conversation. Accordingly, acoustic echo cancellation (AEC) is a widely researched topic.

The traditional approach to AEC is based on the application of adaptive filters~\cite{haensler_acousticechocontrol, Lee_blockbased-filters, shin_NLMS-AP-algos} to estimate the impulse response (IR) of the loudspeaker-enclosure-microphone (LEM) system. Based on the IR, an echo estimate can be computed and subtracted from the microphone signal. While the normalized least mean squared (NLMS) algorithm~\cite{steinert_lowdelayhandsfree} and Kalman filters~\cite{KALMAN, enzner_vary_fdaf, franzen_LowDelayICC_INTERSPEECH} are likely the most prominent solutions, various adaptive filter approaches were continuously improved. Typically a subsequent postfilter is used to suppress residual echo further~\cite{KuechEnzner_StateSpacePartitionedFDAF, franzen_RES_ICASSP}. 

The use of deep neural networks (DNNs) for AEC was proposed just recently, first in the form of residual echo suppression (RES) postfilters (PFs)~\cite{Schwarz_NN_FF_RES, carbajal_RES_ICASSP}. Newer approaches combine AEC with RES and noise suppression in a single DNN~\cite{wang_NN_AEC_18, wang_NN_AEC_19}, treating the echo suppression task as a source separation problem. However, while the reported echo suppression was impressive, it was accompanied by a noticeable decline in the quality of the resulting near-end (NE) speech component. This observation was confirmed in our previous work~\cite{Franzen2021}, in which it was shown that it is the better choice to have a separate AEC DNN to estimate the echo.
	
Based on these findings we propose a two-stage model in which the first Y-Net (AEC DNN) receives the loudspeaker and microphone signals, estimating an echo, while the second \mbox{Y-Net} (PF DNN) receives the microphone signal after echo subtraction and, as a further novelty, the echo estimate output of the first Y-Net, instead of the loudspeaker signal. As the fully convolutional recurrent network (FCRN) structure has proven its strong performance in coded speech enhancement~\cite{zhao_CNN}, in noise suppression~\cite{Strake2019, strake_SingleStage_ICASSP}, including a second-ranked Interspeech 2020 Deep Noise Suppression Challenge proposal~\cite{Strake2020}, and in AEC~\cite{Franzen2021}, we adapt this topology for our Y-Net.

The remainder of this paper is structured as follows: In Section 2, an overview of the two-stage network structure and its individual stages is given. The training and experimental setup is described in Section 3. In Section 4, the experimental results of our model are discussed and compared to the AEC Challenge's baseline approach. Section 5 provides conclusions.

\section{System Overview and Proposed Model}

The speech enhancement in this paper is completely performed by DNNs. The proposed model is based on the findings of \cite{strake_SingleStage_ICASSP} and \cite{Franzen2021}, in which it was already shown that FCRN models perform well separately on AEC and noise suppression. However, it was also found that training a single network for the combined task of AEC and noise suppression leads to significant NE speech distortions. As a result, we propose a two-stage model (Y$^2$-Net) in which the tasks of AEC and noise suppression are performed by consecutive network stages (Y-Nets).

\setlength{\abovecaptionskip}{0.5mm}
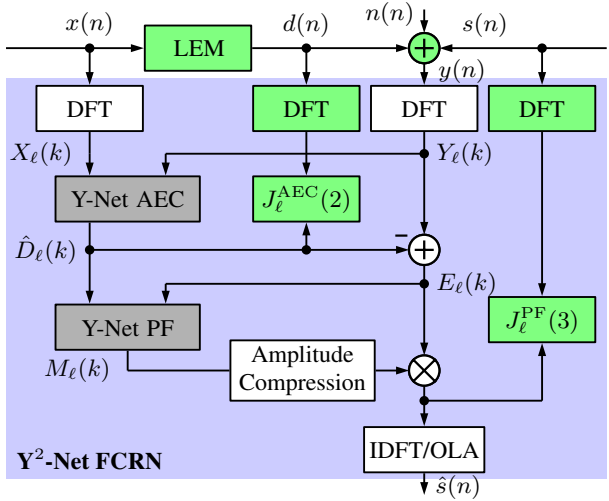
\begin{figure}
	\centering
	\begin{tikzpicture} [scale = 1]
	
	\fill[blue!20] (-1.1,-0.4) rectangle (6.8,-5.705);
	
	\draw (0.0,-5.455) node (lab)		{\textbf{Y$^2$-Net FCRN}};	
	
	\draw (0,0) node 		(x) 		[dspnodefull,label=above:$x(n)$]	{}
	+ (-1.1,0) coordinate	(x1) {}
	++ (1.425,0) 	node 		(IR)		[dspfilter, fill=green!50, text width=2em, minimum height=6mm]	{LEM}
	++ (1.425,0)	node 	(d)			[dspnodefull]							{}
	+  (0,-0.8)	node 		(DFTJ1)		[dspfilter, fill=green!50, text width=3em, minimum height=6mm]	{DFT}
	+ (0,0.3)	node		(dl)									{$d(n)$}
	++ (1.55,0)	node 		(y)			[dspadder]					{}
	+ (0.5,-0.3)	node		(yl)									{$y(n)$}

	+ (0,0.625)	node 		(n)										{}
	+ (-0.45,0.45) node 		(n1)	{$n(n)$}
	
	+ (1.55,0)	node		(s) 		[dspnodefull]	{}
	+ (1.55,-0.8)	node 		(DFTJ2)		[dspfilter, fill=green!50, text width=3em, minimum height=6mm]	{DFT}
	+ (2.4,0)	coordinate		(s2)	{}

	+ (0.8,0.3)	node 		(sl)										{$s(n)$}
	++ (0,-0.8)	node 		(DFT2)		[dspfilter, fill=white, text width=3em, minimum height=6mm]	{DFT}
	++ (0,-0.6)	node 		(P2)		[dspnodefull,label=right:$Y_\ell(k)$]	{}
	+ (-3.4,0)	coordinate	(i2)									{}
	+ (-4.05,0)	coordinate	(i1)									{}
	++ (0,-1.275) node 		(sub)		[dspadder, label={[xshift=-0.3cm, yshift=-0.2cm]\textbf{--}}]	{}
	+ (-4.4,0)	node		(s1)		[dspnodefull,label=left:$\hat{D}_\ell(k)$]				{}
	+ (-3.9,0.675) node 	(Y1)		[dspfilter, text width=6em, minimum height=6mm, fill=black!30]	{Y-Net AEC}
	+ (-1.55,0.675) node 	(J1)		[dspfilter, text width=4em, minimum height=6mm, fill=green!50]	{$J_\ell^\mathrm{AEC}(\ref{eq:J_AEC})$}
	+ (-1.55,0)	node		(j1)		[dspnodefull] {}
	++ (0,-0.45) node 		(P3)		[dspnodefull,label=right:$E_\ell(k)$]	{}
	+ (-3.4,0)	coordinate	(i3)									{}
	++ (0,-1.15)	node 	(mask)		[dspmixer, fill=white]					{}
	
	+ (-1.6,0)	node 	(AC)		[dspfilter,text width=6em, minimum height=8mm, fill=white]	{}
	
	+ (-3.9,0)	coordinate 	(m1)									{}
	+ (-3.9,0)	node 		(m1l)		[label=left:$M_\ell(k)$]		{}
	+ (-3.9,0.575)	node 	(Y2)		[dspfilter,text width=6em, minimum height=6mm, fill=black!30]	{Y-Net PF}
	
	++ (0,-0.4)	node		(m2)		[dspnodefull]			{}
	+ (1.55,1.075) node		(J2)		[dspfilter, text width=4em, minimum height=6mm, fill=green!50]	{$J_\ell^\mathrm{PF}(\ref{eq:J_PF})$}
	+ (1.55,0)	coordinate	(j2)		{}
	
	++ (0,-0.65)	node 		(IDFT)		[dspfilter, fill=white, text width=5em, minimum height=6mm]	{IDFT/OLA}
	++ (0,-0.6)	coordinate 		(r)									{}	
	+  (0.45,0.1) node 		(r1) 	{$\hat{s}(n)$};	
	
	\draw (0,-0.8) node 		(DFT1)		[dspfilter, fill=white, text width=3em, minimum height=6mm]	{DFT}
	+ (0,-0.6)	coordinate		(P1)
	++ (0,-0.6) node 			(P1l)		[label=left:$X_\ell(k)$]	{};
	
	\draw[white, fill=white] (sub) circle (2mm);
	\draw (sub) node [dspadder] (t1) {};
	\draw[white, fill=white] (mask) circle (2mm);
	\draw (mask) node [dspmixer] (t2) {};
	
	\draw[green!50, fill=green!50] (y) circle (2mm);
	\draw (y) node [dspadder] (t3) {};
	
	\draw ([xshift=-5mm] Y1.north) coordinate (Y1c1);
	\draw ([xshift=+5mm] Y1.north) coordinate (Y1c2);
	\draw ([xshift=-5mm] Y1.south) coordinate (Y1c3);
	\draw ([xshift=-5mm] Y2.north) coordinate (Y2c1);
	\draw ([xshift=+5mm] Y2.north) coordinate (Y2c2);
	
	\draw (AC) node [text width=6em, align=center] {Amplitude Compression};
	
	\begin{scope}[start chain]
	
	\chainin (x1);
	\chainin (IR) 	[join=by dspconn];
	\chainin (y) 	[join=by dspconn];
	\chainin (DFT2) [join=by dspconn];
	\chainin (sub) 	[join=by dspconn];
	\chainin (mask) [join=by dspconn];
	\chainin (IDFT) [join=by dspconn];
	\chainin (r) 	[join=by dspconn];
	
	\chainin (n);
	\chainin (y) 	[join=by dspconn];
	\chainin (s2);
	\chainin (s) 	[join=by dspline];
	\chainin (y) 	[join=by dspconn];
	
	\chainin (x);
	\chainin (DFT1) [join=by dspconn];
	\chainin (P1)	[join=by dspline];
	\chainin (Y1c1)	[join=by dspconn];
	
	\chainin (P2);
	\chainin (i2) 	[join=by dspline];
	\chainin (Y1c2)	[join=by dspconn];
	
	\chainin (Y1c3);
	\chainin (s1)	[join=by dspline];
	\chainin (sub)	[join=by dspconn];
	
	\chainin (s1);
	\chainin (Y2c1) [join=by dspconn];
	
	\chainin (P3);
	\chainin (i3) 	[join=by dspline];
	\chainin (Y2c2)	[join=by dspconn];

	\chainin (Y2);
	\chainin (m1)	[join=by dspline];
	\chainin (AC)	[join=by dspline];
	\chainin (mask)	[join=by dspconn];
	
	\chainin (d);
	\chainin (DFTJ1)	[join=by dspconn];
	\chainin (J1)	[join=by dspconn];
	\chainin (j1);
	\chainin (J1)	[join=by dspconn];
	\chainin (s);
	\chainin (DFTJ2)	[join=by dspconn];
	\chainin (J2)	[join=by dspconn];
	\chainin (m2);
	\chainin (j2)	[join=by dspline];
	\chainin (J2)	[join=by dspconn];

	\end{scope}
	\end{tikzpicture}
	\caption{\textbf{Proposed Y$^2$-Net model} (light blue), comprising AEC and PF, embedded into the \textbf{synthetic training setup} (green parts). Details of the Y-Net structure are displayed in Fig. \ref{fig:Submodel}.}
	\label{fig:Model}
\end{figure}

The proposed  Y$^2$-Net model structure for our training setup is depicted in Figure \ref{fig:Model}. The input signals are the loudspeaker reference signal $x(n)$ and the microphone signal \mbox{$y(n)=s(n)+d(n)+n(n)$} with NE speech $s(n)$, noise $n(n)$, and echo $d(n)$, and $n$ being the sample index. For synthetic datasets, $d(n)$ can be computed by applying an LEM model to the reference signal $x(n)$, simulating the effects of loudspeaker distortions and near-end room characteristics.

Our Y$^2$-Net operates entirely in the DFT domain at a sampling rate of 16 kHz. After applying a 1st order high-pass on both $x(n)$ and $y(n)$ to eliminate time-varying biases~\cite{Halimeh2020}, the signals are divided into frames of $N_\mathrm{T}=424$ samples with a frame shift of $50$\% and subject to a square root Hann windowing. The frames are zero-padded so that a $K=512$-point DFT can be applied. The resulting frequency domain representations $X_\ell(k)$ and $Y_\ell(k)$ with frame index $\ell$ and frequency bin index $k \in \mathcal{K}=\{0,1,...,K$\!$-$\!$1\}$ are fed into the first stage of the model (Y-Net AEC). This stage is trained to provide an echo estimate $\hat{D}_\ell(k)$. Unlike other DNN AEC systems (e.g.,~\cite{wang_NN_AEC_18, wang_NN_AEC_19}), but just as classical hands-free systems and also our recent AEC DNN work~\cite{Franzen2021}, $\hat{D}_\ell(k)$ is then subtracted from $Y_\ell(k)$ to obtain the intermediate enhanced signal $E_\ell(k)$. Subsequently, the second model stage (Y-Net PF) performs RES and noise suppression by computing a complex mask $M_\ell(k)$ based on the inputs $E_\ell(k)$ and $\hat{D}_\ell(k)$---a concept so far only known for RES after traditional AEC filters~\cite{carbajal_RES_ICASSP}. We follow the second-ranked non-real-time approach of the Interspeech 2020 Deep Noise Suppression Challenge (with a real-time capable model)~\cite{Strake2020} by employing mask amplitude compression and multiplying its result with $E_\ell(k)$ according to
\vskip-5pt
\begin{equation}
	\hat{S}_\ell(k) = E_\ell(k) \cdot \tanh(|M_\ell(k)|) \cdot\frac{M_\ell(k)}{|M_\ell(k)|}, \label{eq:mask}
\end{equation}
and transform $\hat{S}_\ell(k)$ back into the time-domain by applying a $512$-point IDFT. The resulting frames are cut back to $N_\mathrm{T}$ samples before applying another square root Hann window and reconstructing the final estimated output signal $\hat{s}(n)$ by overlap-add (OLA). The challenge rules require an algorithmic latency defined as $T_1=T+T_s\le40\,\mathrm{ms}$. Our model yields a frame length of $T=26.5\,\mathrm{ms}$ and a frame shift of $T_s=13.25\,\mathrm{ms}$, resulting in $T_1=39.75\,\mathrm{ms}\le40\,\mathrm{ms}$.
\setlength{\abovecaptionskip}{1.5mm}
\begin{figure}
	\centering
	\begin{tikzpicture}
		
	\fill [fill=black!30] (39.5mm, -38.0mm) rectangle (-39.5mm, 37.5mm);
	\fill [yellow!20] (-5mm, -25.0mm) rectangle (-38.5mm, 34.4mm);
	\fill [yellow!20] (5mm, -25.0mm) rectangle (38.5mm, 34.4mm);
		
	\matrix (m1) [row sep= 2.5mm, column sep= 3mm]
	{
		\node[coordinate]								(m00)	{};						&
		\node[coordinate]						(m01)	{};						&
		%			\node[coordinate]						(mX)	{};						&
		\node[coordinate]						(m02)	{};						&
		%			\node[coordinate]						(mX)	{};						&
		\node[coordinate]						(m03)	{};						&
		\node[coordinate]						(m04)	{};				\\
		%			\node[coordinate]						(m04)	{};						\\
		%-----------------------------------------------------------------
		\node[coordinate]						(m10)	{};						&
		\node[coordinate]						(m11)	{};						&
		%			\node[coordinate]						(mX)	{};						&
		\node[coordinate]						(m12)	{};						&
		%			\node[coordinate]						(mX)	{};						&
		\node[coordinate]						(m13)	{};						&
		\node[coordinate]						(m14)	{};	\\
		%-----------------------------------------------------------------
		\node[dspfilter, fill=white, text width=9em, minimum height=2em]		(m20)	{Conv(N$\times$1,F)};	&
		\node[coordinate]						(m21)	{};						&
		%			\node[coordinate]						(mX1)	{};						&
		\node[coordinate]						(m22)	{};						&
		%			\node[coordinate]						(mX2)	{};						&
		\node[coordinate]						(m23)	{};						&
		\node[dspfilter, fill=white, text width=9em, minimum height=2em]		(m24)	{Deconv(N$\times$1,2F)$_{/2}$};	\\
		%-----------------------------------------------------------------
		\node[dspnodefull]						(m30)	{};						&
		\node[coordinate]						(m31)	{};						&
		%			\node[coordinate]						(mX)	{};						&
		\node[coordinate]						(m32)	{};						&
		%			\node[coordinate]						(mX)	{};						&
		\node[coordinate]						(m33)	{};						&
		\node[dspadder, fill=white]						(m34)	{};	\\
		%-----------------------------------------------------------------
		\node[dspfilter, fill=white, text width=9em, minimum height=2em]		(m40)	{Conv(N$\times$1,F)$_{/2}$};	&
		\node[coordinate]						(m41)	{};						&
		%			\node[coordinate]						(mX)	{};						&
		\node[coordinate]						(m42)	{};						&
		%			\node[coordinate]						(mX)	{};						&
		\node[coordinate]						(m43)	{};						&
		\node[dspfilter, fill=white, text width=9em, minimum height=2em]		(m44)	{Deconv(N$\times$1,2F)};	\\
		%-----------------------------------------------------------------
		\node[coordinate]						(m50)	{};						&
		\node[coordinate]						(m51)	{};						&
		%			\node[coordinate]						(mX)	{};						&
		\node[coordinate]						(m52)	{};						&
		%			\node[coordinate]						(mX)	{};						&
		\node[coordinate]						(m53)	{};						&
		\node[coordinate]							(m54)	{};	\\
		%-----------------------------------------------------------------
		\node[dspfilter, fill=white, text width=9em, minimum height=2em]		(m60)	{Conv(N$\times$1,2F)};	&
		\node[coordinate]						(m61)	{};						&
		%			\node[coordinate]						(mX)	{};						&
		\node[coordinate]						(m62)	{};						&
		%			\node[coordinate]						(mX)	{};						&
		\node[coordinate]						(m63)	{};						&
		\node[dspfilter, fill=white, text width=9em, minimum height=2em]		(m64)	{Deconv(N$\times$1,F)$_{/2}$};	\\
		%-----------------------------------------------------------------
		\node[dspnodefull]						(m70)	{};						&
		\node[coordinate]						(m71)	{};						&
		%			\node[coordinate]						(mX)	{};						&
		\node[coordinate]						(m72)	{};						&
		%			\node[coordinate]						(mX)	{};						&
		\node[coordinate]						(m73)	{};						&
		\node[dspadder, fill=white]							(m74)	{};	\\
		%-----------------------------------------------------------------
		\node[dspfilter, fill=white, text width=9em, minimum height=2em]		(m80)	{Conv(N$\times$1,2F)$_{/2}$};	&
		\node[coordinate]						(m81)	{};						&
		%			\node[coordinate]						(mX)	{};						&
		\node[coordinate]						(m82)	{};						&
		%			\node[coordinate]						(mX)	{};						&
		\node[coordinate]						(m83)	{};						&
		\node[dspfilter, fill=white, text width=9em, minimum height=2em]		(m84)	{Deconv(N$\times$1,F)};	\\
		%-----------------------------------------------------------------
		\node[coordinate]						(m90)	{};						&
		\node[coordinate]						(m91)	{};						&
		%			\node[coordinate]						(mX5)	{};						&
		\node[coordinate]						(m92)	{};						&
		%			\node[coordinate]						(mX6)	{};						&
		\node[coordinate]						(m93)	{};						&
		\node[coordinate]						(m94)	{};	\\
		%-----------------------------------------------------------------
		\\
		\node[dspfilter, fill=white, text width=9em, minimum height=2em]		(m100)	{ConvLSTM(N$\times$1,F)}; &
		\node[coordinate]						(m101)	{};						&
		\node[coordinate]						(m102)	{};	&
		\node[coordinate]						(m103)	{};						&
		\node[dspfilter, fill=white, text width=9em, minimum height=2em]		(m104)	{Conv(N$\times$1,C)};	\\
		%-----------------------------------------------------------------
		\node[coordinate]						(m110)	{};						&
		\node[coordinate]						(m111)	{};						&
		\node[coordinate]						(m112)	{};						&
		\node[coordinate]						(m113)	{};						&
		\node[coordinate]						(m114)	{};	\\
	};

	\draw ([xshift=-51.2mm, yshift=-0.6mm] m114) node (lab) {\textbf{Y-Net (EF)}};
	\draw ([xshift=-5.5mm, yshift=-1mm] m114) node (W) {${W}_\ell(k)$};

	\draw[white, fill=white] (m34) circle (2mm);
	\draw (m34) node [dspadder] (t1) {};
	\draw[white, fill=white] (m74) circle (2mm);
	\draw (m74) node [dspadder] (t2) {};

	\draw ([xshift=-3mm] m71) coordinate (m71a);
	\draw ([xshift=-3mm] m31) coordinate (m31a);
	\draw ([xshift=+3mm] m73) coordinate (m73a);
	\draw ([xshift=+3mm] m33) coordinate (m33a);
	
	\foreach \i [evaluate = \i as \j using int(\i+1),
	evaluate = \i as \k using int(\i+2),] in {2,4,6}
	{
		\begin{scope}[start chain]
		\chainin (m\i0);
		\chainin (m\k0) [join=by dspconn];
		\end{scope}
		\begin{scope}[start chain]
		\chainin (m\i4);
		\chainin (m\j4) [join=by dspline];
		\chainin (m\k4) [join=by dspconn];
		\end{scope}
	}S
			
	\begin{scope}[start chain]
		\chainin (m80);
		\chainin (m100) [join=by dspconn];
		\chainin (m110) [join=by dspline];
		\chainin (m112) [join=by dspline];
		\chainin (m12) [join=by dspline];
		\chainin (m14) [join=by dspline];
		\chainin (m24) [join=by dspconn];
		
		\chainin (m30);
		\chainin (m31a) [join=by dspline];		
		\chainin (m73a) [join=by dspline];
		\chainin (m74) [join=by dspconn];
		
		\chainin (m70);
		\chainin (m71a) [join=by dspline];
		\chainin (m33a) [join=by dspline];
		\chainin (m34) [join=by dspconn];
		
		\chainin (m24);
		\chainin (m34) [join=by dspconn];
		\chainin (m64);
		\chainin (m74) [join=by dspconn];
		
		\chainin (m84);
		\chainin (m104) [join=by dspconn];
	\end{scope}
	
	\draw ([yshift=-2mm] m114) coordinate (end);
	\draw[dspconn] (m104) -> (end);
	
	\draw[dspconn,<-] ([xshift=+3mm] m20.north) --node[left,text=black!100, yshift=0mm] {} +(0,+2.2em);
	\draw ([xshift=+7mm, yshift=0.2mm] m00) node {${V}_\ell(k)$};
	
	\draw[dspconn,<-] ([xshift=-2mm] m20.north) --node[right,text=black!100, yshift=0mm] {} +(0,+2.2em);
	\draw ([xshift=-6mm, yshift=0.2mm] m00) node {${U}_\ell(k)$};
	
	\draw ([xshift=-10.5mm, yshift=-1mm] m10) node {M$\times 1 \times$2C$_\text{in}$};
	\draw ([xshift=-9mm] m30) node {M$\times 1 \times$F};
	\draw ([xshift=-9.5mm] m50) node {\nicefrac{M}{2}$\times 1 \times$F};
	\draw ([xshift=-10mm] m70) node {\nicefrac{M}{2}$\times 1 \times$2F};
	\draw ([xshift=-10mm] m90) node {\nicefrac{M}{4}$\times 1 \times$2F};
	
	\draw ([xshift=9.5mm] m14) node {\nicefrac{M}{4}$\times 1 \times$F};
	\draw ([xshift=10mm] m34) node {\nicefrac{M}{2}$\times 1 \times$2F};
	\draw ([xshift=10mm] m54) node {\nicefrac{M}{2}$\times 1 \times$2F};
	\draw ([xshift=9mm] m74) node {M$\times1 \times$F};
	\draw ([xshift=9mm] m94) node {M$\times1 \times$F};
	\draw ([xshift=9mm, yshift=0mm] m114) node {M$\times1 \times$C$_\text{out}$};
	
	\draw ([yshift=-4.5mm, xshift=-5mm] m80.south east) node {Encoder};
	
	\draw ([yshift=-4.5mm, xshift=5mm] m84.south west) node {Decoder};
	
	\end{tikzpicture}
	
	\caption{\textbf{Proposed Y-Net FCRN structure} (here with EF) used for each stage of the proposed Y$^2$-Net model (see Figure 1) with DFT-domain inputs ${U}_\ell(k)$ and ${V}_\ell(k)$, and output ${W}_\ell(k)$.}
	\label{fig:Submodel}
\end{figure}
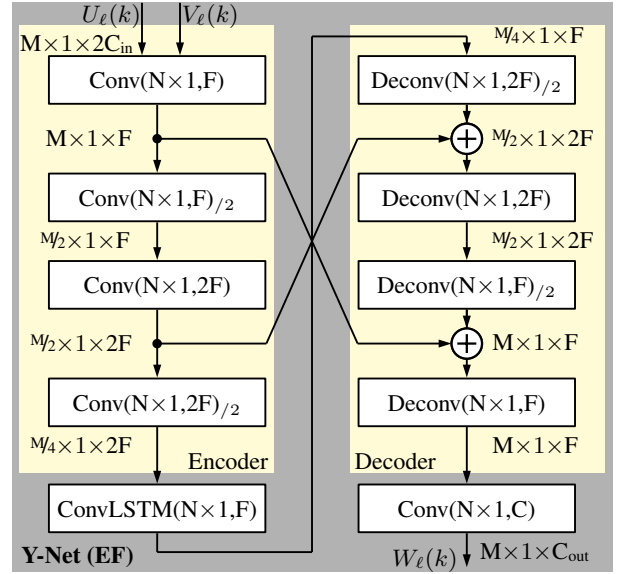

The topology of the Y-Net FCRN is depicted in Figure \ref{fig:Submodel}. The feature dimensions of each layer are denoted as \mbox{\textit{feature axis $\times$ time axis $\times$ number of feature maps}}. The kernel size is $N\times1$ with $N=24$ in feature direction and $1$ in time direction. Determining the number of filter kernels throughout the network layers, we choose \mbox{$F=70$}.

The Y-Net FCRN uses an encoder-decoder structure with two generic inputs ${U}_\ell(k)$ and ${V}_\ell(k)$ and one output ${W}_\ell(k)$ (hence the name Y-Net). The encoder features four convolutional layers. Every other layer is configured with a stride of $2$ in the feature axis (marked as $_{/2}$), thereby halving the feature axis each time. The first two layers use $F$ kernels, while the last two layers use $2F$ kernels. The decoder is designed inversely to the encoder, using strides on deconvolutional layers~\cite{DECONV} to restore the original feature axis length. All convolutional and deconvolutional layers use a leaky ReLU activation function~\cite{leakyReLU}. The bottleneck between encoder and decoder consists of a convolutional LSTM~\cite{ConvLSTM, strake_SingleStage_ICASSP} layer with $F$ kernels, a tanh() activation function, and the "hard sigmoid" recurrent activation function as defined by {\tt TensorFlow2}~\cite{Abadi2016}. Two skip connections are employed. The convolutional output layer after the decoder block yields ${W}_\ell(k)$ and uses $C$ kernels and a linear activation. 

Although the specific input and output features for each stage are selected individually, both stages' inputs and outputs consist of spectral frequency bins. These bins are divided into two channels each ($C=2$), containing their real and imaginary components. By concatenation along the channel axis, $2C=4$ real-valued input channels are obtained. To preserve the feature axis size $M$, it must be chosen to be dividable by $4$. With a $512$-point DFT yielding $257$ non-redundant spectral frequency bins, the frequency dimension is set to $M=260$ and channels are zero-padded accordingly. 

Note that we investigate the Y-Net AEC either having identical early fusion (EF) as shown in Figure \ref{fig:Submodel}, dubbing the model as Y$^2$-Net EF/EF, or an AEC variant called late fusion (LF), where separate encoder paths are used for the inputs \mbox{${U}_\ell(k)={X}_\ell(k)$} and ${V}_\ell(k)={Y}_\ell(k)$. Each encoder path of LF models has the same layers and parameters as the single path of EF, and both paths are concatenated just before the ConvLSTM block. The skip connections then originate from the encoder of ${Y}_\ell(k)$. The corresponding entire model with EF-type PF (Fig. \ref{fig:Submodel}) is labeled Y$^2$-Net LF/EF. The effects of EF and LF are investigated in more detail in \cite{Franzen2021}. As a Y-Net with LF has considerably more parameters to train than with EF, we choose $F=60$ for the Y-Net AEC of Y$^2$-Net LF/EF.

\section{Experimental Setup}

\subsection{Dataset}

The model is trained with synthetic files from the database provided by Microsoft for the Interspeech 2021 AEC Challenge~\cite{Cutler2021}, in the following labeled as $\mathcal{D}_\mathrm{syn}$. The audio files were created with varying conditions, including single-talk, double-talk, both NE and FE noise, as well as simulated nonlinear loudspeaker distortions. The dataset consists of $10,\!000$ audio samples of $10$\,s length each. All speaker audio was taken from the {\tt LibriVox} project~\cite{LibriVox} by picking a $10$\,s audio sample of a randomly selected speaker for the FE signal. The NE signal is then chosen from a different speaker and contains \mbox{$3$\,s-$7$\,s} of audio padded to the length of the FE signal. The IRs for the FE echoes are generated using Microsoft's {\tt Project Triton}~\cite{ProjectTriton} technology with reverberation times $T_{60}$ between $0.2$\,s and $1.2$\,s. For $80$\% of the files the FE signal is also subject to one of various nonlinear functions before applying the IR to simulate loudspeaker distortions. A more in-depth description can be found in \cite{Sridhar2021}. A number of $500$ files of $\mathcal{D}_\mathrm{syn}$ were created with unseen speakers and IRs, and are used as preliminary synthetic testset $\mathcal{D}_\mathrm{syn}^\mathrm{test,pre}$ for instrumental measurements. The remaining files are split into $8,\!000$ files ($\mathcal{D}_\mathrm{syn}^\mathrm{train}$) for training and $1,\!500$ for validation between epochs ($\mathcal{D}_\mathrm{syn}^\mathrm{val}$).

For each audio sample, the dataset contains the noisy microphone signal $y(n)$ and the reference signal $x(n)$, as well as the speech components for NE speech $s(n)$ and echo $d(n)$ at the microphone. The NE noise $n(n)$ in the provided data is only present during the duration of NE speech, which we find as not representative for real scenarios. Therefore, we augmented the dataset with custom noise by adding an additional NE noise component randomly taken from the database provided by Microsoft for the Interspeech 2021 Deep Noise Suppression Challenge~\cite{Reddy2021} with an SNR between $0$\,dB and $40$\,dB. As NE speech and FE echo were not altered, the SER of audio files in $\mathcal{D}_\mathrm{syn}$ lies between -$10$\,dB and $10$\,dB as originally designed by the dataset creators. The FE noise also remains unchanged.

Furthermore, the performance of the network is evaluated on two real test datasets, both being provided by the challenge organizers. The preliminary real testset $\mathcal{D}_\mathrm{real}^\mathrm{test,pre}$ and the final blind test set $\mathcal{D}_\mathrm{real}^\mathrm{test}$ each consist of 800 real-world recordings comprising double-talk, FE single-talk, or NE single-talk.

\subsection{Training Targets and Loss Computation}

\textit{The stages of the model are trained with different targets according to the assigned task.} The Y-Net AEC computes an echo estimate $\hat{D}_\ell(k)$, which is then subtracted from the microphone signal to yield an enhanced intermediate signal $E_\ell(k)$. The Y-Net PF, in turn, provides a complex mask $M_\ell(k)$, which is applied to $E_\ell(k)$ according to (\ref{eq:mask}), yielding the final estimated enhanced signal $\hat{S}_\ell(k)$. The training is conducted using a standard MSE loss, which is computed for the individual network stages as
\vskip-12pt
\begin{align}
	J_{\ell}^\mathrm{AEC} & = \frac{1}{K} \sum_{k \in \mathcal{K}} \big| \hat{D}_\ell(k) - D_\ell(k) \big|^2 \label{eq:J_AEC} \\
	J_{\ell}^\mathrm{PF} & = \frac{1}{K} \sum_{k \in \mathcal{K}} \big| \hat{S}_\ell(k) - S_\ell(k) \big|^2. \label{eq:J_PF}
\end{align}

The network is trained in two steps: First, the Y-Net AEC is pretrained separately using its individual loss (\ref{eq:J_AEC}) on $\mathcal{D}_\mathrm{syn}^\mathrm{train}$. Then, both Y-Net AEC and Y-Net PF are jointly trained on $\mathcal{D}_\mathrm{syn}^\mathrm{train}$ with the pretrained weights for the Y-Net AEC and a weighted loss function
\begin{equation}
	J_\ell  = \alpha J_{\ell}^\mathrm{AEC} + (1-\alpha) J_{\ell}^\mathrm{PF} \label{eq:J}
\end{equation}
with $\alpha=0.25$. This combined loss for both stages ensures that the AEC performance of the first stage is tuned to yield a residual echo characteristic which can be efficiently suppressed by the Y-Net PF. {\it Note that (\ref{eq:J}) is a fundamental difference of our work to earlier source separation methods for AEC~\cite{wang_NN_AEC_18, wang_NN_AEC_19}, as we strongly suggest an intermediate echo estimate $\hat{D}_\ell(k)$ being used in losses (\ref{eq:J_AEC}) and (\ref{eq:J}).}

\subsection{Training Parameters}

During all stages of the training, the Adam optimizer~\cite{adam_optimizer} is used in its standard parameter configuration. The batch size is set to $16$ and the backpropagation-through-time~\cite{BPTT} (BPTT) unrolling sequence length per batch is set to $50$ frames. The initial learning rate is set to $5 \cdot 10^{-3}$. For every $3$ epochs without loss improvement on $\mathcal{D}_\mathrm{syn}^\mathrm{val}$, the learning rate is multiplied with a factor of $0.6$. The training is stopped after $100$ epochs, if the validation loss does not improve for 10 consecutive epochs, or if the learning rate drops below $5 \cdot 10^{-4}$.

\section{Results and Discussion}

We use three metrics to measure the performance on our preliminary synthetic testset $\mathcal{D}_\mathrm{syn}^\mathrm{test,pre}$. Some of these metrics require individual enhanced signal components for NE speech $\tilde{s}(n)$, echo $\tilde{d}(n)$, and noise $\tilde{n}(n)$, which are derived from the enhanced signal $\hat{s}(n) = \tilde{s}(n) + \tilde{d}(n) + \tilde{n}(n)$ and from $s(n)$, $d(n)$, and $n(n)$, using the black-box signal separation approach according to ITU-T Recommendation P.1110 \cite{ITU_P1110}, with more details in \cite{fingscheidt_signalseparation, fingscheidt_blackbox, steinert_instrumentaldistortionassessment}. All metrics using components obtained by this approach are marked with an index \rm{BB}.

\setlength{\abovecaptionskip}{2mm}
\begin{table*}[t]
	\caption{\textbf{Instrumental ratings} on the \textbf{preliminary synthetic testset} $\mathcal{D}_\mathrm{syn}^\mathrm{test,pre}$, and real-time factor (RTF). Y-Net AEC does not perform noise suppression and is therefore separated from the other models. Best results per table segment are marked in \textbf{bold}, second-best results are \underline{underlined}. Our model proposed for the Interspeech 2021 AEC Challenge is marked in \textbf{bold}.}
	\label{tab:synthetic}
	\centering
	\setlength\tabcolsep{3.5pt}
	\renewcommand{\arraystretch}{1.00}
	\begin{tabular}{ p{42mm}p{2.5mm}p{10mm}<{\centering}p{10mm}<{\centering}p{10mm}<{\centering}p{10mm}<{\centering}p{2.5mm}p{9mm}<{\centering}p{2.5mm}p{9mm}<{\centering}p{2.5mm}p{9mm}<{\centering}p{2.5mm}p{7mm}<{\centering}}
		\toprule
		\multicolumn{1}{c}{Input (microphone) signals:} && \multicolumn{4}{c}{full mixture y(n)} & &\multicolumn{1}{c}{$d(n)$} && \multicolumn{1}{c}{$n(n)$} && \multicolumn{1}{c}{$s(n)$} &&\\\cmidrule{1-1}\cmidrule{3-6}\cmidrule{8-8}\cmidrule{10-10}\cmidrule(){12-12}
		\multicolumn{1}{c}{Method:   /   \textbf{Metric}:} && \multicolumn{1}{c}{\textbf{PESQ}} & \multicolumn{1}{c}{\textbf{ERLE$_\text{BB}$}} & \multicolumn{1}{c}{\boldmath{$\Delta$}\textbf{SNR$_\text{BB}$}} & \multicolumn{1}{c}{\textbf{PESQ$_\text{BB}$}} && \multicolumn{1}{c}{\textbf{ERLE}} && \multicolumn{1}{c}{\boldmath{$\Delta$}\textbf{SNR}}&& \multicolumn{1}{c}{\textbf{PESQ}} && \multicolumn{1}{c}{\textbf{RTF}}\\
		\midrule
		Y-Net AEC (EF), pretraining		&& 2.14 & \textbf{15.86} & 2.77 & \textbf{3.74} && \textbf{26.36} && \textbf{0.86} && \textbf{4.51} && 0.42\\
		Y-Net AEC (EF), joint training	&& \textbf{2.16} & 15.10 & \textbf{2.78} & {3.73} && 24.20 && 0.85 && {4.48} && 0.42\\
		\midrule
		Baseline && 1.95 & 27.57 & 5.31 & \textbf{3.31} && 45.34 && 1.11 && \textbf{4.61} && 0.05 \\
		Y$^2$-Net (EF/EF) ({E}, {X}) && \underline{2.46} & \textbf{32.17} & 10.31 & 3.09 && 52.19 && 26.04 && \underline{4.44} && 0.95 \\
		Y$^2$-Net (EF/EF) ({E}, $\hat{\mathrm{D}}$) && 2.40 & 31.55 & 10.28 & 3.12 && 51.98 && \textbf{29.27} && 4.35 && 0.95 \\
		Y$^2$-Net (LF/EF) ({E}, {X}) && \underline{2.46} & 31.72 & \textbf{10.69} & \underline{3.21} && \textbf{52.45} && 28.58 && 4.31 && 0.96 \\
		\textbf{Y$^2$-Net (LF/EF) ({E}, $\hat{\textbf{D}}$)} && \textbf{2.47} & \underline{32.13} & {10.60} & 3.12 && \underline{52.35} && \underline{29.10} && 4.30 && 0.96 \\
		{Y-Net (EF), loss (3) only (as \cite{wang_NN_AEC_19})} && 2.40 & 27.50 & \underline{10.62} & 3.11 && 38.77 && 28.83 && 4.09 && 0.97 \\
		\bottomrule
	\end{tabular}
\vspace{-1.5mm}
\end{table*}

\begin{table}[t]
	\caption{\textbf{Instrumental DECMOS ratings} on the \textbf{preliminary real testset} $\mathcal{D}_\mathrm{real}^\mathrm{test,pre}$.}
	\label{tab:real}
	\centering
	\setlength\tabcolsep{1.5pt}
	\renewcommand{\arraystretch}{1.00}
	\begin{tabular}{ p{30mm}p{8mm}<{\centering}p{8mm}<{\centering}p{8mm}<{\centering}p{8mm}<{\centering}p{8mm}<{\centering}}
		\toprule
		& \multicolumn{2}{c}{\textbf{single-talk}} & \multicolumn{2}{c}{\textbf{double-talk}}&\\\cmidrule(lr){4-5}\cmidrule(lr){2-3}
		\multicolumn{1}{c}{\textbf{Method}} & \multicolumn{1}{c}{\textbf{FE}} & \multicolumn{1}{c}{\textbf{NE}} & \multicolumn{1}{c}{\textbf{FE}} & \multicolumn{1}{c}{\textbf{NE}}& \multicolumn{1}{c}{\textbf{total}}\\
		\midrule
		Baseline & 3.63 & \textbf{3.64} & 3.01 & 2.78 & 3.27 \\
		Y$^2$-Net (EF/EF) ({E}, X)  & 3.65 & 3.55 & 3.59 & 3.13 & 3.48 \\
		Y$^2$-Net (EF/EF) ({E}, $\hat{\mathrm{D}}$)  & 3.52 & 3.58 & 3.54 & 3.19 & 3.46 \\
		Y$^2$-Net (LF/EF) ({E}, X)  & \textbf{3.66} & 3.63 & 3.52 & 3.14 & 3.49 \\
		\textbf{Y$^2$-Net (LF/EF) ({E}, $\hat{\textbf{D}}$)}  & \textbf{3.66} & {3.63} & \textbf{3.61} & \textbf{3.21} & \textbf{3.53} \\
		\bottomrule
	\end{tabular}
\vspace{-1.5mm}
\end{table}

The quality of the NE speech component is evaluated on the wideband PESQ MOS LQO~\cite{ITU_P862.2, ITU_P862.2_Corr}. The echo reduction is measured by the echo return loss enhancement \mbox{$\text{ERLE}(n) = 10 \cdot \log \big(d^2(n) / \tilde{d}^2(n) \big)\text{ [dB].}$}
After applying a first order IIR smoothing filter with factor $0.99$, the final ERLE value is computed as an average over the entire dataset. The noise reduction performance of a model is measured as the SNR difference between the signals $y(n)$ and $\hat{s}(n)$ in [dB] and is defined as
$\Delta\text{SNR} = 10\cdot\log_{10}\left(\sum_{n}\tilde{s}^2(n)/\sum_{n}\tilde{n}^2(n)\right)$ $-
10\cdot\log_{10}\left(\sum_{n}s^2(n)/\sum_{n}n^2(n)\right)$
with sample index $n$ spanning the entire utterance. To give a clear overview of the models' performance, the tests are conducted under several conditions. The four center columns of Table \ref{tab:synthetic} marked as \textit{full mixture} provide results for the preliminary testset $\mathcal{D}_\mathrm{syn}^\mathrm{test,pre}$. PESQ scores are evaluated on the full processed audio file $\hat{s}(n)$ but also on $\tilde{s}(n)$, the latter labeled as PESQ$_\mathrm{BB}$. The three metrics in the columns to the right of the full mixture columns represent special conditions under which only the specified signal component is used as microphone signal $y(n)$. These conditions observe the models' performance beyond double-talk scenarios: How well does the model handle echo or noise if no NE speech is present? And can it pass through clean NE speech without distortions? This metric setup in turn also requires to set $x(n)=0$ (unless $y(n) = d(n)$) as no echo is present in this case. For these metrics the respective enhanced signal component is equal to the entire enhanced signal, so no black-box approach is needed to compute them. For $\Delta$SNR under the condition of $y(n) = n(n)$, $\Delta\text{SNR}$ is simplified to $\Delta\text{SNR} = 10\log_{10}\left(\sum_{n}n^2(n)/\sum_{n}\tilde{n}^2(n)\right)$.

The challenge organizers conducted a subjective evaluation of our models on the final test dataset $\mathcal{D}_\mathrm{real}^\mathrm{test}$ based on the ITU-T P.808 framework~\cite{ITU-P808}. Four scenarios were evaluated: FE single-talk "echo DMOS" test (P.831~\cite{ITU-P831}), NE single-talk "MOS" test (P.808), double-talk "echo DMOS" test (P.831), and double-talk "other degradation DMOS" test (P.831). For double-talk, the "echo DMOS" test will be labeled as "FE" in Tables \ref{tab:real} and \ref{tab:real2}, describing the echo annoyance. The "other degradation DMOS" test is dubbed as "NE", describing quality of NE speech. More details can be found in \cite{Sridhar2021}. 

The challenge organizers also provided an instrumental measurement algorithm DECMOS, which yields MOS values that are claimed to be highly correlated to those of the described subjective listening tests. This is important for the evaluation of our proposed models, as our other instrumental metrics cannot be applied to real datasets because the separate signal components $s(n)$, $d(n)$ and $n(n)$ are not available.

Table \ref{tab:synthetic} shows the results of our proposed model on the preliminary synthetic testset $\mathcal{D}_\mathrm{syn}^\mathrm{test,pre}$. 
In the first segment of the table, we report our AEC Y-Net (EF) after pre-training, and after joint training, confirming that joint training hardly compromises echo reduction performance at the output of the first Y-Net AEC. In the second table segment, our Y$^2$-Net \mbox{models} are tested with two variants of Y-Net PF inputs: the common approach of using intermediate output and reference signal (E, X), and the new approach using intermediate output and estimated echo ({E}, $\hat{\mathrm{D}}$) as depicted in Figure \ref{fig:Model}. It can be seen that our Y$^2$-Net model outperforms the challenge baseline significantly in terms of ERLE, SNR and full mixture PESQ. The baseline can score better values for PESQ$_\mathrm{BB}$ due to not performing noise reduction. As the LF/EF Y$^2$-Net with ({E}, $\hat{\mathrm{D}}$) shows four 1$^\text{st}$ and 2$^\text{nd}$ ranks among the Y$^2$-Net variants, and $s(n)$-only PESQ metrics of $\ge 4.30$ points hardly yield audible differences, we select this model as our proposed model (marked in bold). To emphasize the advantage of a two-stage approach, the performance of a single-stage Y-Net, which employs enhanced signal estimation by AEC and noise suppression in one step (similar to \cite{wang_NN_AEC_19}), is shown in the last line. {The number of feature maps $F$ is set to $100$} for this model to achieve a complexity comparable to our Y$^2$-Net approaches. We observe that our Y$^2$-Net with intermediate echo estimation not only allows for a considerably better echo suppression, but also delivers better overall PESQ.

\begin{table}[t]
	\caption{\textbf{Subjective ratings} according to ITU-T P.808 and \textbf{objective ratings} (DECMOS) on the \textbf{(blind) real testset} $\mathcal{D}_\mathrm{real}^\mathrm{test}$.}
	\label{tab:real2}
	\centering
	\setlength\tabcolsep{1pt}
	\renewcommand{\arraystretch}{1.00}
	\begin{tabular}{ p{10mm}<{\centering}p{30mm}p{8.5mm}<{\centering}p{8.5mm}<{\centering}p{8.5mm}<{\centering}p{8.5mm}<{\centering}}
		\toprule
		&& \multicolumn{2}{c}{\textbf{single-talk}} & \multicolumn{2}{c}{\textbf{double-talk}}\\\cmidrule(lr){5-6}\cmidrule(lr){3-4}
		Metric&\multicolumn{1}{c}{\textbf{Method}} & \multicolumn{1}{c}{\textbf{FE}} & \multicolumn{1}{c}{\textbf{NE}} & \multicolumn{1}{c}{\textbf{FE}} & \multicolumn{1}{c}{\textbf{NE}}\\
		\midrule
		subj&Baseline & \textbf{3.82} & \textbf{4.18} & \textbf{4.04} & 3.45 \\
		subj&Y$^2$-Net (LF/EF) (\textbf{E}, $\hat{\textbf{D}}$)  & 3.73 & 4.16 & 3.72 & \textbf{3.53} \\
		\midrule
		obj&Baseline & 3.48 & 3.72 & 3.02 & 2.78 \\
		obj&Y$^2$-Net (LF/EF) (\textbf{E}, $\hat{\textbf{D}}$)  & \textbf{3.49} & \textbf{3.76} & \textbf{3.55} & \textbf{3.18} \\
		\bottomrule
	\end{tabular}
\vspace{-1.0mm}
\end{table}

Table \ref{tab:synthetic} also reports real-time factors (RTFs), which represent the relative computing time of a time frame compared to the frame shift. Values below $1.0$ imply that the inference with this model can be performed in real-time. The RTF was measured on an {\tt Intel i7-10510U} quad-core CPU at $1.8$\,GHz. All RTFs are below $1.0$, therefore the models comply with the challenge rules for the real-time track.

Table \ref{tab:real} shows the results of the proposed model on the preliminary real testset $\mathcal{D}_\mathrm{real}^\mathrm{test,pre}$ for the DECMOS metric. Our Y$^2$-Net models perform better than the baseline, with especially notable improvements in the double-talk scenario. Among the tested variants, the Y$^2$-Net (LF/LE) ({E}, $\hat{\mathrm{D}}$) maintains the best performance in total and was therefore selected as final model.

As reported in Table \ref{tab:real}, we find that both blind real testset $\mathcal{D}_\mathrm{real}^\mathrm{test}$ single-talk conditions in DECMOS metrics and subjective listening MOS yield similar results for baseline and \mbox{Y$^2$-Net} (LF/LE) ({E}, $\hat{\mathrm{D}}$) in Table \ref{tab:real2} as well. For double-talk FE, the strong DECMOS score is in line with Table \ref{tab:real}. In contrast, the subjective challenge test results show an unexpected higher score of the baseline, while the NE double-talk condition again shows higher scores for our proposed Y$^2$-Net, subjective and objective. Note that our model shows 0.46 DECMOS points improvement over the baseline in double-talk. Our informal listening supports the DECMOS metric in all conditions.

\section{Conclusions}

In this work we present a novel Y$^2$-Net, a two-stage fully convolutional recurrent network model for acoustic echo cancellation and noise suppression, with an echo estimate as output of the first stage. As a further novelty for DNN AEC, we propose to use the estimated echo signal of the first stage as part of the second stage's input, as well as a joint training for both stages still including the intermediate echo estimate loss term. Compared to the baseline approach, the model significantly increases performance during double-talk by an average of $0.46$ MOS points in the DECMOS metric.

\bibliographystyle{IEEEtran}

\bibliography{biblio}

\begin{thebibliography}{10}
\providecommand{\url}[1]{#1}
\csname url@samestyle\endcsname
\providecommand{\newblock}{\relax}
\providecommand{\bibinfo}[2]{#2}
\providecommand{\BIBentrySTDinterwordspacing}{\spaceskip=0pt\relax}
\providecommand{\BIBentryALTinterwordstretchfactor}{4}
\providecommand{\BIBentryALTinterwordspacing}{\spaceskip=\fontdimen2\font plus
\BIBentryALTinterwordstretchfactor\fontdimen3\font minus
  \fontdimen4\font\relax}
\providecommand{\BIBforeignlanguage}[2]{{%
\expandafter\ifx\csname l@#1\endcsname\relax
\typeout{** WARNING: IEEEtran.bst: No hyphenation pattern has been}%
\typeout{** loaded for the language `#1'. Using the pattern for}%
\typeout{** the default language instead.}%
\else
\language=\csname l@#1\endcsname
\fi
#2}}
\providecommand{\BIBdecl}{\relax}
\BIBdecl

\bibitem{haensler_acousticechocontrol}
E.~H\"ansler and G.~Schmidt, \emph{{Acoustic Echo and Noise Control: A
  Practical Approach}}.\hskip 1em plus 0.5em minus 0.4em\relax Hoboken, NJ,
  USA: Wiley-Interscience, 2004.

\bibitem{Lee_blockbased-filters}
J.~Lee and C.~Un, ``{Block Realization of Multirate Adaptive Digital
  Filters},'' \emph{IEEE Transactions on Acoustics, Speech, and Signal
  Processing}, vol.~34, no.~1, pp. 105--117, Feb. 1986.

\bibitem{shin_NLMS-AP-algos}
H.~Shin, A.~H. Sayed, and W.~Song, ``{Variable Step-Size NLMS and Affine
  Projection Algorithms},'' \emph{IEEE Signal Processing Letters}, vol.~11,
  no.~2, pp. 132--135, Feb. 2004.

\bibitem{steinert_lowdelayhandsfree}
K.~Steinert, M.~Sch\"onle, C.~Beaugeant, and T.~Fingscheidt, ``{Hands-free
  System with Low-Delay Subband Acoustic Echo Control and Noise Reduction},''
  in \emph{{Proc. of ICASSP}}, {Las Vegas, NV, USA}, Apr. 2008, pp. 1521--1524.

\bibitem{KALMAN}
R.~E. Kalman, ``{A New Approach to Linear Filtering and Prediction Problems:
  Journal of Basic Engineering},'' \emph{Journal of Basic Engineering},
  vol.~82, no.~1, pp. 35--45, Mar. 1960.

\bibitem{enzner_vary_fdaf}
G.~Enzner and P.~Vary, ``{Frequency-Domain Adaptive Kalman Filter for Acoustic
  Echo Control in Hands-Free Telephones},'' \emph{{S}ignal {P}rocessing
  (Elsevier)}, vol.~86, no.~6, pp. 1140--1156, Jun. 2006.

\bibitem{franzen_LowDelayICC_INTERSPEECH}
J.~Franzen and T.~Fingscheidt, ``{A Delay-Flexible Stereo Acoustic Echo
  Cancellation for DFT-Based In-Car Communication (ICC) Systems},'' in
  \emph{{Proc.\ of INTERSPEECH}}, Stockholm, Sweden, Aug. 2017, pp. 181--185.

\bibitem{KuechEnzner_StateSpacePartitionedFDAF}
F.~Kuech, E.~Mabande, and G.~Enzner, ``{State-Space Architecture of the
  Partitioned-Block-Based Acoustic Echo Controller},'' in \emph{{Proc.\ of
  ICASSP}}, Florence, Italy, May 2014, pp. 1295--1299.

\bibitem{franzen_RES_ICASSP}
J.~Franzen and T.~Fingscheidt, ``{An Efficient Residual Echo Suppression for
  Multi-Channel Acoustic Echo Cancellation Based on the Frequency-Domain
  Adaptive Kalman Filter},'' in \emph{Proc.\ of ICASSP}, Calgary, AB, Canada,
  Apr. 2018, pp. 226--230.

\bibitem{Schwarz_NN_FF_RES}
A.~Schwarz, C.~Hofmann, and W.~Kellermann, ``{Spectral Feature-Based Nonlinear
  Residual Echo Suppression},'' in \emph{Proc.\ of WASPAA}, New Paltz, NY, USA,
  Oct. 2013, pp. 1--4.

\bibitem{carbajal_RES_ICASSP}
G.~Carbajal, R.~Serizel, E.~Vincent, and E.~Humbert, ``{Multiple-Input Neural
  Network-Based Residual Echo Suppression},'' in \emph{Proc.\ of ICASSP},
  Calgary, AB, Canada, Apr. 2018, pp. 231--235.

\bibitem{wang_NN_AEC_18}
H.~Zhang and D.~Wang, ``{Deep Learning for Acoustic Echo Cancellation in Noisy
  and Double-Talk Scenarios},'' in \emph{Proc.\ of INTERSPEECH}, Hyderabad,
  India, Sep. 2018, pp. 3239--3243.

\bibitem{wang_NN_AEC_19}
H.~Zhang, K.~Tan, and D.~Wang, ``{Deep Learning for Joint Acoustic Echo and
  Noise Cancellation with Nonlinear Distortions},'' in \emph{Proc.\ of
  INTERSPEECH}, Graz, Austria, Sep. 2019, pp. 4255--4259.

\bibitem{Franzen2021}
J.~Franzen, E.~Seidel, and T.~Fing\-scheidt, ``{AEC in a Netshell: On Target
  and Topology Choices for FCRN Acoustic Echo Cancellation},'' in \emph{Proc.
  of ICASSP}, Toronto, Canada, Jun. 2021, pp. 156--160.

\bibitem{zhao_CNN}
Z.~Zhao, H.~Liu, and T.~Fingscheidt, ``{Convolutional Neural Networks to
  Enhance Coded Speech},'' \emph{{IEEE/ACM Trans. on Audio, Speech, and
  Language Processing}}, vol.~27, no.~4, pp. 663--678, Apr. 2019.

\bibitem{Strake2019}
M.~Strake, B.~Defraene, K.~Fluyt, W.~Tirry, and T.~Fing\-scheidt, ``{Separated
  Noise Suppression and Speech Restoration: LSTM-Based Speech Enhancement in
  Two Stages},'' in \emph{Proc. of WASPAA}, New Paltz, NY, USA, Oct. 2019, pp.
  239--243.

\bibitem{strake_SingleStage_ICASSP}
M.~Strake, B.~Defraene, K.~Fluyt, W.~Tirry, and T.~Fing\-scheidt, ``{Fully
  Convolutional Recurrent Networks for Speech Enhancement},'' in \emph{Proc.\
  of ICASSP}, Barcelona, Spain, May 2020, pp. 6674--6678.

\bibitem{Strake2020}
M.~Strake, B.~Defraene, K.~Fluyt, W.~Tirry, and T.~Fing\-scheidt,
  ``{INTERSPEECH 2020 Deep Noise Suppression Challenge: A Fully Convolutional
  Recurrent Network (FCRN) for Joint Dereverberation and Denoising},'' in
  \emph{Proc. of INTERSPEECH}, Shanghai, China, Oct. 2020, pp. 2467--2471.

\bibitem{Halimeh2020}
M.~Halimeh, T.~Haubner, A.~Briegleb, A.~Schmidt, and W.~Kellermann,
  ``{Combining Adaptive Filtering and Complex-Valued Deep Postfiltering for
  Acoustic Echo Cancellation},'' \emph{ResearchGate:
  10.13140/RG.2.2.14083.94241}, Nov. 2020.

\bibitem{DECONV}
V.~Dumoulin and F.~Visin, ``{A Guide to Convolution Arithmetic for Deep
  Learning},'' \emph{arXiv:1603.07285}, Jan. 2018.

\bibitem{leakyReLU}
A.~L. Maas, A.~Y. Hannun, and A.~Y. Ng, ``{Rectifier Nonlinearities Improve
  Neural Network Acoustic Models},'' in \emph{Proc.\ of ICML Workshop on Deep
  Learning for Audio, Speech, and Language Processing}, Atlanta, GA, USA, Jun.
  2013, pp. 1--6.

\bibitem{ConvLSTM}
X.~Shi, Z.~Chen, H.~Wang, D.-Y. Yeung, W.~Wong, and W.~Woo, ``{Convolutional
  LSTM Network: A Machine Learning Approach for Precipitation Nowcasting},'' in
  \emph{Proc.\ of NIPS}, Montreal, QC, Canada, Dec. 2015, pp. 802--810.

\bibitem{Abadi2016}
M.~Abadi \emph{et~al.}, ``{TensorFlow: Large-Scale Machine Learning on
  Heterogeneous Distributed Systems},'' \emph{arXiv:1603.04467}, Mar. 2016.

\bibitem{Cutler2021}
R.~Cutler, A.~Saabas, T.~Parnamaa, M.~Loide, S.~Sootla, M.~Purin, H.~Gamper,
  S.~Braun, K.~Sorensen, R.~Aichner, and S.~Srinivasan, ``{INTERSPEECH 2021
  Acoustic Echo Cancellation Challenge: Datasets and Testing Framework},'' in
  \emph{Proc. of INTERSPEECH \emph{(accepted for publication)}}, Brno, Czech
  Republic, Sep. 2021.

\bibitem{LibriVox}
\BIBentryALTinterwordspacing
H.~McGuire, ``{LibriVox: Acoustical liberation of books in the public
  domain},'' 2005--present, [accessed 26.03.2021]. [Online]. Available:
  \url{https://librivox.org/}
\BIBentrySTDinterwordspacing

\bibitem{ProjectTriton}
\BIBentryALTinterwordspacing
N.~Raghuvanshi and J.~Snyder, ``{Project Triton},'' 2010--present, [accessed
  26.03.2021]. [Online]. Available:
  \url{https://www.microsoft.com/en-us/research/project/project-triton/}
\BIBentrySTDinterwordspacing

\bibitem{Sridhar2021}
K.~Sridhar, R.~Cutler, A.~Saabas, T.~Parnamaa, M.~Loide, H.~Gamper, S.~Braun,
  R.~Aichner, and S.~Srinivasan, ``{ICASSP 2021 Acoustic Echo Cancellation
  Challenge: Datasets, Testing Framework, and Results},''
  \emph{arXiv:2009.04972}, Oct. 2020.

\bibitem{Reddy2021}
C.~{K A Reddy}, H.~Dubey, K.~Koishida, A.~Nair, V.~Gopal, R.~Cutler, S.~Braun,
  H.~Gamper, R.~Aichner, and S.~Srinivasan, ``{INTERSPEECH 2021 Deep Noise
  Suppression Challenge},'' \emph{arXiv:2101.01902}, Jan. 2021.

\bibitem{adam_optimizer}
D.~P. Kingma and J.~Ba, ``{Adam: A Method for Stochastic Optimization},'' in
  \emph{Proc.\ of ICLR}, San Diego, CA, USA, May 2015, pp. 1--15.

\bibitem{BPTT}
R.~Pascanu, T.~Mikolov, and Y.~Bengio, ``{On the Difficulty of Training
  Recurrent Neural Networks},'' in \emph{Proc. of ICML}, Atlanta, GA, USA, Jun.
  2013, p. 1310–1318.

\bibitem{ITU_P1110}
``{ITU-T Recommendation P.1110, Wideband Hands-Free Communication in Motor
  Vehicles},'' ITU-T, Mar. 2017.

\bibitem{fingscheidt_signalseparation}
T.~Fingscheidt and S.~Suhadi, ``{Quality Assessment of Speech Enhancement
  Systems by Separation of Enhanced Speech, Noise, and Echo},'' in
  \emph{{Proc.\ of INTERSPEECH}}, {Antwerp, Belgium}, Aug. 2007, pp. 818--821.

\bibitem{fingscheidt_blackbox}
T.~Fingscheidt, S.~Suhadi, and K.~Steinert, ``{Towards Objective Quality
  Assessment of Speech Enhancement Systems in a Black Box Approach},'' in
  \emph{{Proc. of ICASSP}}, {Las Vegas, NV, USA}, Apr. 2008, pp. 273--276.

\bibitem{steinert_instrumentaldistortionassessment}
K.~Steinert, S.~Suhadi, T.~Fingscheidt, and M.~Sch\"onle, ``{Instrumental
  Speech Distortion Assessment of Black Box Speech Enhancement Systems},'' in
  \emph{Proc.\ of IWAENC}, {Seattle, WA, USA}, Sep. 2008, pp. {273--276}.

\bibitem{ITU_P862.2}
``{ITU-T Recommendation P.862.2, Wideband Extension to Recommendation P.862 for
  the Assessment of Wideband Telephone Networks and Speech Codecs},'' ITU-T,
  Nov. 2007.

\bibitem{ITU_P862.2_Corr}
``{ITU-T Recommendation P.862.2 Corrigendum 1, Wideband Extension to
  Recommendation P.862 for the Assessment of Wideband Telephone Networks and
  Speech Codecs},'' ITU-T, Oct. 2017.

\bibitem{ITU-P808}
``{ITU-T Recommendation P.808, Subjective Evaluation of Speech Quality with a
  Crowdsourcing Approach},'' ITU-T, Jun. 2018.

\bibitem{ITU-P831}
``{ITU-T Recommendation P.831, Subjective Performance Evaluation of Network
  Echo Cancellers},'' ITU-T, Dec. 1998.

\end{thebibliography}

\end{document}